\documentclass[reprint, aps,pra, showpacs]{revtex4-1}
\usepackage{dcolumn}  
\usepackage{bm}       

\usepackage{amsmath,amssymb, graphicx}
\usepackage{amsthm}
\usepackage{color}

\newtheorem{definitionenv}{Definition}
\newtheorem{lemmaenv}[definitionenv]{Lemma}
\newtheorem{theoremenv}[definitionenv]{Theorem}
\newtheorem{corollaryenv}[definitionenv]{Corollary}
\newtheorem{propositionenv}[definitionenv]{Proposition}
\newtheorem{conjectureenv}[definitionenv]{Conjecture}
\newtheorem{exampleenv}{Example}
\newtheorem{app-lemmaenv}[section]{Lemma}
\newtheorem{remarkenv}[definitionenv]{Remark}

\newenvironment{definition}{\begin{definitionenv}\rm}{\end{definitionenv}}
\newenvironment{lemma}{\begin{lemmaenv}\rm}{\end{lemmaenv}}
\newenvironment{theorem}{\begin{theoremenv}\rm}{\end{theoremenv}}
\newenvironment{corollary}{\begin{corollaryenv}\rm}{\end{corollaryenv}}
\newenvironment{example}{\begin{exampleenv}\rm}{\end{exampleenv}}
\newenvironment{proposition}{\begin{propositionenv}\rm}{\end{propositionenv}}
\newenvironment{conjecture}{\begin{conjectureenv}\rm}{\end{conjectureenv}}
\newenvironment{app-lemma}{\begin{app-lemmaenv}\rm}{\end{app-lemmaenv}}
\newenvironment{remark}{\begin{remarkenv}\rm}{\end{remarkenv}}

\newcommand{\bd}{\begin{definition}}
\newcommand{\ed}{\end{definition}}
\newcommand{\edp}{\hspace*{\fill} $\Box$
                 \end{definition}}
\newcommand{\bl}{\begin{lemma}}
\newcommand{\el}{\end{lemma}}
\newcommand{\elp}{\hspace*{\fill} $\Box$
                 \end{lemma}}
\newcommand{\bt}{\begin{theorem}}
\newcommand{\et}{\end{theorem}}
\newcommand{\etp}{\hspace*{\fill} $\Box$
                 \end{theorem}}
\newcommand{\bc}{\begin{corollary}}
\newcommand{\ec}{\end{corollary}}
\newcommand{\ecp}{\hspace*{\fill} $\Box$
                 \end{corollary}}
\newcommand{\bcj}{\begin{conjecture}}
\newcommand{\ecj}{\end{conjecture}}
\newcommand{\be}{\begin{example}}
\newcommand{\ee}{\end{example}}
\newcommand{\eep}{\hspace*{\fill} $\Box$
                 \end{example}}
\newcommand{\bp}{\begin{proposition}}
\newcommand{\ep}{\end{proposition}}
\newcommand{\epp}{\hspace*{\fill} $\Box$
                 \end{proposition}}
\newcommand{\br}{\begin{remark}}
\newcommand{\er}{\end{remark}}
\newcommand{\erp}{\hspace*{\fill} $\Box$
                 \end{remark}}

\newcommand{\bPf}[1]{\begin{proof} #1 \end{proof}}

\DeclareMathOperator{\rank}{rank}
\DeclareMathOperator{\Row}{Row}

\newcommand{\ket}[1]{|#1\rangle}

\newcommand{\ketbra}[2]{|#1\rangle\langle#2|}

\newcommand{\beq}[1]{\begin{equation*}{ #1 }\end{equation*}}
\newcommand{\beql}[2]{\begin{equation}\label{#1}{ #2 }\end{equation}}
\newcommand{\beqs}[1]{\begin{align*} #1 \end{align*}}
\newcommand{\beqsl}[2]{\begin{equation}\label{#1}\begin{aligned} #2 \end{aligned}\end{equation}}

\newcommand{\bcase}[1]{\begin{cases} #1 \end{cases}}

\newcommand{\bsmtx}[1]{\left[ \begin{smallmatrix}  #1  \end{smallmatrix} \right]}

\newcommand{\fa} {~\forall~}

\newcommand{\eq}[1]{\eqref{#1}}       

\newcommand{\ZZ}{{\mathbb Z}}

\newcommand{\hE}{{\hat E}}
\newcommand{\hH}{{\hat H}}

\newcommand{\sC}{{\cal C}}
\newcommand{\sE}{{\cal E}}
\newcommand{\sG}{{\cal G}}

\newcommand{\sK}{{\cal K}}

\newcommand{\sN}{{\cal N}}
\newcommand{\sS}{{\cal S}}
\newcommand{\af}{\alpha}

\newcommand{\veps}{\varepsilon}
\newcommand{\la}{\lambda}

\newcommand{\sig}{\sigma}
\newcommand{\vph}{\varphi}
\newcommand{\La}{\Lambda}

\newcommand{\dg}{\dagger}

\newcommand{\gen}[1]{\langle {#1} \rangle}
\newcommand{\teq}{\triangleq}

\newcommand{\floor}[1]{\left\lfloor {#1} \right\rfloor}

\begin{document}


\title{
  On the Hardnesses of Several Quantum Decoding Problems
}
\thanks{
Part of this paper was presented in
the 2012 International Symposium on Information Theory and its Applications (ISITA 2012),
Hawaii, USA, October 28--31, 2012.
}



\author{Kao-Yueh Kuo}
 \email{d9761808@oz.nthu.edu.tw}

\author{Chung-Chin Lu}
 \email{cclu@ee.nthu.edu.tw}
 \altaffiliation[]{Chung-Chin Lu is the person to correspond with.}

\affiliation{National Tsing Hua University, Hsinchu 30013, Taiwan}


\begin{abstract}
We classify the time complexities
of three important decoding problems
for quantum stabilizer codes.
First, regardless of the channel model,
quantum bounded distance decoding
is shown to be NP-hard,
like what Berlekamp, McEliece and Tilborg did
for classical binary linear codes in 1978.
Then over the depolarizing channel,
the decoding problems
for finding a most likely error
and for minimizing the decoding error probability
are also shown to be NP-hard.
Our results indicate
that finding a polynomial-time decoding algorithm for general
stabilizer codes may be impossible, but this, on the other hand,
strengthens the foundation of quantum code-based cryptography.
\end{abstract}


\maketitle


\section{Introduction} \label{sec:introduction}
For classical binary linear codes,
Berlekamp, McEliece and Tilborg \cite{BMT78} considered
the classical bounded distance decoding (CBDD),
and asked a simpler decision problem \textsc{Coset Weights}.
They proved that \textsc{Coset Weights} is NP-complete
so that CBDD is NP-hard.
In the theory of computation \cite{Sipser06},
a decision (yes-or-no) problem
is NP-complete if it is in NP and all NP problems are
reducible to it in polynomial time.
A computational problem (not necessary a decision problem)
is NP-hard if
an NP-complete problem
is reducible to it in polynomial time.
The fact that CBDD is NP-hard indicates that
it may be impossible to find a polynomial-time decoding algorithm
for general classical binary linear codes.
But McEliece then pointed out that the result actually provides
a foundation of code-based cryptography \cite{McEliece78}.
The importance of code-based cryptography has grown
recently because code-based cryptosystems
appear to have strong resistance against the attacks performed
by quantum computers \cite{BBD09}, \cite{BLP10}. 

It is known that quantum stabilizer codes
can be related to classical self-orthogonal codes \cite{CRSS97}, \cite{CRSS98}, \cite{GotPhd}.
But Poulin and Chung \cite{PC08} pointed out that
decoding stabilizer codes
could be very different from decoding classical codes,
since rather than finding a most likely error,
finding a most likely error coset is desired.
Later, Hsieh and Gall \cite{HG10} showed that,
over a special Pauli channel 
regarding {\it Hamming weight} as the weight metric,
the quantum decoding problems are NP-hard,
no matter whether a most likely error
or a most likely error coset is desired.
More recently, Fujita \cite{Fujita12} showed that,
regarding the {\it generalized weight} as the weight metric,
a bounded distance decoding for stabilizer codes,
as a decision problem, is NP-complete.
The generalized weight is an important metric since it is usually used to define
the minimum distance of stabilizer codes \cite{CRSS98}, \cite{GotPhd}, \cite{CEL99}
to directly reflect the error-correction capability in number of qubits.
And when stabilizer codes are used over the depolarizing channel,
a direct extension of the classical binary-symmetric channel \cite{BS98},
the generalized weight can be used to determine the probability of an error
(as in Eq.~\eq{eq_Ensem}, Sec.~\ref{DecDpCh}).
However, the complexity (hardness) of an optimal decoding
over the depolarizing channel,
i.e., finding a most likely error coset under the generalized weight,
is still unknown in the literature.

In this paper, we classify the hardnesses of several quantum decoding problems
including the aforementioned optimal decoding problem.
In the beginning,
regarding the generalized weight but regardless of any specific channel model,
quantum bounded distance decoding (QBDD) is considered.
Fujita considered a similar problem (see Lemma 2 of \cite{Fujita12})
without restricting the check matrix to be of full row rank.
We will take this restriction and show that QBDD is NP-hard.
Then over the depolarizing channel,
quantum maximum likelihood decoding (QMLD)
and quantum minimum-error-probability decoding (QMEPD)
are considered,
where the first is to find a most likely error
and the second is to find a most likely error coset.
Assisted by the NP-hardness of QBDD,
we will show that both QMLD and QMPED are also NP-hard.

The paper is organized as follows.
In Section \ref{StbCode},
the foundation of stabilizer codes is reviewed
and the required notations are defined.
In Section \ref{BdDec},
QBDD is considered
and shown to be NP-hard.
In Section~\ref{DecDpCh},
QMLD and QMEPD over the depolarizing channel
are considered and both shown to be NP-hard.
In Section~\ref{Summary}, a conclusion is given.

\section{Stabilizer Codes} \label{StbCode}

In this section,
we define the state space we work with and stabilizer codes.
The stabilizer codes will be related
to even-length classical binary codes
under the symplectic inner product
and the generalized weight
\cite{CRSS97}, \cite{CRSS98}, \cite{GotPhd}, \cite{CEL99}, \cite{NC00}, \cite{KL10}.

Let $V_1$ be the state space of one qubit,
which is a 2-dimensional complex inner product space
spanned by an orthonormal computational basis $\{ \ket{0},\ket{1} \}$.
Let
$\sG_1 \teq \{ \pm I, \pm iI, \pm X, \pm iX, \pm Y, \pm iY, \pm Z, \pm iZ \}$
be the Pauli Goup on $V_1$,
where
$$
I \teq \bsmtx{1 &0\\ 0 &1}, ~
X \teq \bsmtx{0 &1\\ 1 &0}, ~
Z \teq \bsmtx{1 &0\\ 0 &-1}, ~
Y \teq \bsmtx{0 &-i\\ i &0} = iXZ.
$$
Then $\sG_n \teq \sG_1^{\otimes n}$ is the Pauli group
on the state space $V \teq V_1^{\otimes n}$ of $n$ qubits.
It is known that two elements in $\sG_n$ either commute or anti-commute.
For each $g\in\sG_n$,
it has a tensor product representation
$$
g = i^{m_0} \sig_1 \otimes \sig_2 \otimes \cdots \otimes \sig_n,
$$
where $m_0 \in \{0,1,2,3\}$ and $\sig_j \in \{I,X,Y,Z\}$ for all $j$.
Let $w(g)$ be the {\it weight} of $g$,
which is defined as the number of non-identity terms
in the tensor product representation of $g$.
For example, $g= i(X\otimes I\otimes Y\otimes Z)\in\sG_4$ has $w(g)=3$.
Since an error possibly affecting a state in $V$
can be written as a linear combination of the elements in $\sG_n$,
according to the error discretization theorem
(Theorem 10.2 of \cite{NC00}),
we can correct a state of $n$ qubits
having errors in $\le t$ qubits
if (and only if)
all error patterns $E\in \sG_n$ of weight $w(E)\le t$ are correctable.
It is known that such an error-correction capability can
be achieved by an $[[n,k,d]]$ stabilizer code with $d\ge 2t+1$ \cite{CRSS98}, \cite{GotPhd}.

Stabilizer codes are defined in the following manner.
Let $\sS$ be a subgroup of $\sG_n$ such that $-I\notin\sS$.
Then $\sS$ is abelian and can be generated by
a set of $n-k$ independent generators as
\beql{StbGp}{
\sS = \gen{g_1,g_2,\cdots,g_{n-k}}
}
for some integer $k\in [0,n]$.
The subgroup $\sS$ has a fixed subspace $\sC(\sS)$ in $V$ defined as
$$
\sC(\sS) \teq \{ \ket{\psi}\in V ~\big|~ g\ket{\psi}=\ket{\psi} \fa g\in \sS \},
$$
which has dimension $2^k$.
The subspace $\sC(\sS)$ is called an $[[n,k]]$ {\it stabilizer code}
with a {\it stabilizer group} $\sS$.
Most properties of $\sC(\sS)$ can be studied through $\sS$.
The {\it normalizer} $\sN(\sS)$ of $\sS$ in $\sG_n$ is defined as
\beq{
\sN(\sS) \teq \{ h\in\sG_n \mid hg=gh \fa g\in\sS \}.
}
Let $\sK \teq \{ \pm I, \pm iI \}$ and
$\sS\sK \teq \sS\vee\sK = \{ \pm g, \pm ig \mid g\in \sS \}$.
Then it is known that
the {\it minimum distance} $d$ of the stabilizer code $\sC(\sS)$
can be defined as
\beql{d}{
d \teq \min\{ w(g) \mid g\in \sN(\sS)\setminus \sS\sK \}.
}

Now we relate stabilizer codes to classical binary linear codes.
Let $\vph: \sG_n \to \ZZ_2^{2n}$ be a group epimorphism
defined by
\beqs{
\varphi(g) &= \varphi(i^{m_0} \sig_1 \otimes \sig_2 \otimes \cdots \otimes \sig_n) \\
&= (x_1 x_2 \cdots x_n | z_1 z_2 \cdots z_n) = (x|z)
}
for all $g\in\sG_n$ under the mapping:
\beq{
\begin{array}{c|cccc}
\sig_j & I & X & Y & Z \\
\hline
x_j & 0 & 1 & 1 & 0 \\
z_j & 0 & 0 & 1 & 1 \\
\end{array}.
}
Let $g,h$ be any two elements in $\sG_n$.
The epimorphism $\vph$ gives us the relation
$\vph(gh) = \vph(g)+\vph(h)$ of group operations.
Define
$$
\La\teq\left[
  {O_{n \times n} \atop I_{n \times n}} \vrule ~
  {I_{n \times n} \atop O_{n \times n}} \right].
$$
Then we have $gh=hg$ iff $\vph(g)\La\vph(h)^T = 0$,
i.e., $\vph(g)$ and $\vph(h)$ are orthogonal
with respect to (w.r.t.)~the symplectic inner product \cite{CRSS98}.
Use the $n-k$ generators of $\sS$ in \eq{StbGp}
to define an $(n-k)\times 2n$ binary matrix
\beql{ChkMtx}{
H\teq
\bsmtx{
  \vph(g_1)\\[-4pt]
  \vdots\\
  \vph(g_{n-k})
  }_{(n-k)\times 2n}.
}
Since $g_i$'s are independent generators of $\sS$,
the rows $\varphi(g_i)$'s in $H$ are linear independent
so that $H$ is of full row rank.
Also since $\sS$ is abelian, we have $H\La H^T=O$
so that the row space $\Row(H)$ of $H$
is a classical binary linear code
$C=\vph(\sS\sK)\subseteq \ZZ_2^{2n}$
which is self-orthogonal w.r.t.~the symplectic inner product.
$H$ is called a {\it check matrix} of the stabilizer code $\sC(\sS)$.
Now for each $(x|z) \in \ZZ_2^{2n}$,
define the {\it generalized weight} of $(x|z)$ as
\beq{
gw(x|z) \teq w_H(x)+w_H(z)-w_H(xz),
}
where $w_H(x)$ is the Hamming weight of $x$,
and $xz$ is the bitwise AND of $x$ and $z$.
A property of the generalized weight is
\beql{gw_ge_wH}{
gw(x|z) \ge \max\{w_H(x), w_H(z)\}
}
and note that
$w(g) = gw(\vph(g))$ for all $g\in\sG_n$.
%
%
%
%
Let
$C^\perp \teq \{ v\in\ZZ_2^{2n} ~|~ v\La H^T = {\bf 0}\}$
be the symplectic dual of $C$.
Then $\vph^{-1}(C^\perp)= \sN(\sS)$,
and the minimum distance $d$ of $\sC(\sS)$ in \eq{d}
can be evaluated by
\beql{dmin}{
d = \min\{ gw(v) ~|~ v\in C^\perp\setminus C \}.
}

\section{Quantum Bounded Distance Decoding} \label{BdDec}

In this section,
we will review the syndrome measurement of stabilizer codes,
and then define the quantum bounded distance decoding (QBDD) problem.
We will consider the constraint that the check matrix in QBDD is of full row rank.
Without this constraint,
Fujita \cite{Fujita12} considered this problem as a decision problem,
and proved that it is NP-complete.
Then he used this fact as a foundation to propose stabilizer code-based cryptosystems.
To further strengthen this foundation,
we will prove that QBDD is NP-hard in this section.
The NP-hardness of QBDD will then help us to classify the hardnesses
of the decoding problems in the next section.

Now we briefly review the syndrome measurement of stabilizer codes
(see Section 10.5 of \cite{NC00} for details).
Consider an $[[n,k]]$ stabilizer code $\sC(\sS)$ as in Sec.~\ref{StbCode}.
Assume that an uncoded state of $k$ qubits is encoded
to a coded state $\ket{\psi}\in \sC(\sS)$ of $n$ qubits.
Let $\rho \teq \ketbra{\psi}{\psi}$ be the channel input
and $\sE(\rho) = E\rho E^\dg$ be the channel output,
provided that the error is some $E\in\sG_n$.
%
To perform the error detection,
the $n-k$ generators of $\sS$ are used to form
$n-k$ syndrome measurements
$$\{(I+g_i)/2, (I-g_i)/2\},~~ i=1,2,\cdots,n-k.$$
For each generator $g_i$ of $\sS$, we have either
$Eg_i=g_iE$ with measurement output being $+1$
or ${Eg_i=-g_iE}$ with measurement output being $-1$,
while the post-measurement state
remains unchanged as $E\rho E^\dg$ with probability one.
Assume that the measurements are performed and the results form an $n-k$ tuple
$ \beta =(\beta_1,\beta_2,\cdots,\beta_{n-k}) \in \{+1,-1\}^{n-k} $.
Map $\beta$ to a binary vector $s=(s_1,s_2,\cdots,s_{n-k})\in\ZZ_2^{n-k}$
by $s_i=0$ if $\beta_i=+1$ and $s_i=1$ if $\beta_i=-1$ for all $i$.
By the discussion in Sec.~\ref{StbCode}, we have $E g_i= g_i E$ iff $\vph(E)\La\vph(g_i)^T=0$.
Let $H$ be defined as in \eq{ChkMtx}.
Then we have $s = \vph(E)\La H^T$, i.e.,
$s$ can be regarded as
a classical binary syndrome vector
generated by the error vector $\vph(E)$
and the check matrix $H$ under the symplectic inner product.

The above discussion suggests a bounded distance decoding problem that,
given a check matrix $H_{(n-k)\times 2n}$ and a syndrome vector $s\in\ZZ_2^{n-k}$,
we need to find an error vector $e\in\ZZ_2^{2n}$
such that $e\La H^T = s$ and $gw(e)\le t$ for some integer $t\ge 0$.
This is a classical computational problem.
Once a solution $e$ is found, up to a global phase,
any $\hE\in\vph^{-1}(e)$
can be an error-correction operator
applying to the post-measurement state $E\rho E^\dg$.
Now we define the decoding problem of finding such an $e$ as
\\[5pt]
{\bf \textsc{Quantum Bounded Distance Decoding} (QBDD) }\\[2pt]
\textsc{Input:}  A full row-rank $m\times 2n$ binary matrix $H$ satisfying $H\La H^T=O$,
                 a binary vector $s\in\ZZ_2^m$,
                 and an integer $t\ge 0$.\\[1pt]
\textsc{Output:} A binary vector $e\in\ZZ_2^{2n}$ satisfying $gw(e)\le t$
                 and $e\La H^T = s$,
                 or a failure indication if such an $e$ does not exist.
\\[8pt]
%
We remark that even if the channel error $E\in\sG_n$ has a weight $w(E)\le t$
and the check matrix $H$ corresponds to a stabilizer code with a minimum distance $d\ge 2t+1$,
there may still exist more than one solution $e$ to QBDD.
This phenomenon is different from the classical decoding due to the evaluation of $d$ in \eq{dmin}.
However, all these solutions will correspond
to the same error-correction effect.
To see that,
first recall $d=\min\{gw(u) ~|~ u\in C^\perp\setminus C\}$, where $C=\Row(H)=\vph(\sS\sK)$.
And consider the {\it stabilizer property} that
for all $g\in\sG_n$,
\beql{StbProperty}{
g\rho g^\dg = \rho \iff g\in\sS\sK,
}
where 
$\sS\sK = \{\pm g, \pm ig \,|\, g\in\sS\}$.
So if $d\ge 2t+1$, $w(E)\le t$, and QBDD has multiple solutions,
then any two solutions $e_1$ and $e_2$ will
satisfy $e_2=e_1+v$ for some $v\in C=\vph(\sS\sK)$,
i.e., $e_1$ and $e_2$ correspond to two errors
$E_1\in\vph^{-1}(e_1)$ and $E_2\in\vph^{-1}(e_2)$
satisfying $E_2 = E_1 g$ for some $g\in\sS\sK$. 
So
\beqsl{SameCorr}{
& E_1^\dg (E\rho E^\dg) E_1
= g E_2^\dg (E\rho E^\dg) E_2 g^\dg \\
&= (E_2^\dg E) g\rho g^\dg (E^\dg E_2)
= E_2^\dg (E\rho E^\dg) E_2 
}
by \eq{StbProperty}, no matter $g$ and $E_2^\dg E$ commute or anti-commute.
Thus $E_1^\dg$ and $E_2^\dg$ has the same error-correction effect.
The phenomenon that
distinct operators can result in
the same error-correction effect
is sometimes called the \textit{degeneracy property} \cite{PC08}, \cite{HG10}.
Whether a stabilizer code has the degeneracy property
depends on how the set of correctable error patterns is defined.
For bounded distance decoding,
the set of correctable error patterns is usually defined as
$\{E\in\sG_n \mid w(E)\le \floor{ \frac{d-1}{2} } \}$.
In this case, a stabilizer code $\sC(\sS)$ is degenerate if its minimum distance
$d=\min\{w(g) \mid g\in \sN(\sS) \setminus \sS\sK\}$
equals $d' \teq \min\{w(g) \mid g\in \sN(\sS)\}$.

The hardness of QBDD reflects the hardness of attacking
a quantum code-based cryptosystem like the one in \cite{Fujita12}
when this kind of system is protected through bounded artificial noise
like McEliece's suggestion \cite{McEliece78}.
To classify the hardness of QBDD, we at first
deal with a decision (yes-or-no) problem:
\\[5pt]
{\bf \textsc{Coset Generalized Weights (CGW)} }\\[2pt]
\textsc{Input:}    An $m\times 2n$ binary matrix $H$ satisfying $H\La H^T=O$,
                   a binary vector $s\in\ZZ_2^m$,
                   and an integer $t\ge 0$.\\[1pt]
\textsc{Question:} There exists a binary vector $e\in\ZZ_2^{2n}$ satisfying $gw(e)\le t$
                   and $e\La H^T = s$.
\\[8pt]
%
Fujita also considered this decision problem and showed that it
is NP-complete in Lemma 2 of \cite{Fujita12}.
For convenience, we restate the result
as
\bt \label{CGW_NpComp}
CGW is NP-complete,
even if $H=[H_X|O]$ or $H=[O|H_Z]$.
\et
%
%
In order to know the hardness of QBDD,
we need the constraint that the check matrix is of full row rank.
So we restrict CGW as follows.
If $H=[O|H_Z]$ is assumed in {CGW},
we say that {CGW} becomes a restricted problem {CGW$_Z$}.
If $H=[O|H_Z]$ is further assumed to have full row rank, we say that
{CGW$_Z$} becomes a further restricted problem {CGW$_{ZF}$}.
We define the two problems for clarity: 
\\[8pt]
{\bf \textsc{CGW$_Z$} }\\[2pt]
\textsc{Input:} An $m\times 2n$ binary matrix $H=[O|H_Z]$,
                   a binary vector $s\in\ZZ_2^m$,
                   and an integer $t\ge 0$.\\[1pt]
\textsc{Question:} There exists a binary vector $e\in\ZZ_2^{2n}$ satisfying $gw(e)\le t$
                   and $e\La H^T = s$.
\\[0pt]
%
%
\\[0pt]
{\bf \textsc{CGW$_{ZF}$} }\\[2pt]
\textsc{Input:} An $m'\times 2n$ binary matrix $H'=[O|H_Z']$
                   having full row rank,
                   a binary vector $s'\in\ZZ_2^{m'}$,
                   and an integer $t'\ge 0$.\\[1pt]
\textsc{Question:} There exists a binary vector $e\in\ZZ_2^{2n}$ satisfying $gw(e)\le t'$
                   and $e\La H'^T = s'$.
\\[8pt]
Note that an $H=[O|H_Z]$ already satisfies $H\La H^T=O$.
The problem {CGW$_Z$} is already NP-complete, as stated in Theorem~\ref{CGW_NpComp}.
Similar to a remark in \cite{BMT78}, the NP-completeness of {CGW$_Z$}
will imply the NP-completeness of {CGW$_{ZF}$}.
We prove this statement rigorously in
%
\bt \label{CGW_ZF_NpComp}
{CGW$_Z$} is polynomial-time reducible to {CGW$_{ZF}$},
and hence {CGW$_{ZF}$} is also NP-complete.
\et
\bPf{
Suppose we have a polynomial-time algorithm for {CGW$_{ZF}$}.
Given an instance of {CGW$_Z$} with some
$H=[O|H_Z]_{m\times 2n}$,
$s\in\ZZ_2^{m}$, and $t\ge 0$,
assume $\rank(H)=m'$.
Then $m'\le m$.
We also have $m'\le n$ from $H\La H^T=O$.
Let $H_Z'$ in {CGW$_{ZF}$}
consist of $m'$ rows of $H_Z$
so that $\Row(H') = \Row(H)$.
Constructing such an $H_Z'$
can be done in polynomial time, as a remark in Sec.~IV of \cite{BMT78}.
Once this is done, then $H = RH'$
for some $m\times m'$ binary matrix $R$.
The matrix $R$ can also be constructed in polynomial time.
(For example, first put $H'$ in a reduced row echelon form $\hH' = EH'$,
which can be done in polynomial time.
Then it is trivial to find a unique $\hat R$ satisfying
$H = \hat R \hH'= \hat R E H'$, which implies $R=\hat R E$.)
Assume that rows $j_1,j_2,\cdots,j_{m'}$ of $H_Z$
compose the rows of $H_Z'$.
By the given $s=(s_1~s_2~\cdots~s_m)$,
let $s'=(s_{j_1}~ s_{j_2}~ \cdots~ s_{j_{m'}})$.
%
Then any $u\in\ZZ_2^{2n}$ satisfying
$u\La H^T=s$ will satisfy $u\La H'^T=s'$.
And if such a $u$ exists,
we must have $s'R^T = u\La H'^T R^T = u\La H^T = s$.
Thus, if the equality $s' R^T = s$ does not hold,
then the answer to {CGW$_{Z}$} is negative.
If the equality $s'R^T = s$ holds,
then let $t'=t$ and
use the polynomial-time algorithm to solve {CGW$_{ZF}$}.
If the answer to {CGW$_{ZF}$} is negative,
then the answer to {CGW$_{Z}$} is also negative
by our construction of $H'_Z$.
Conversely, if the answer to {CGW$_{ZF}$} is positive,
then there exists a vector $e\in\ZZ^{2n}$
satisfying $gw(e)\le t$ and $e\La H'^T = s'$.
Then this $e$ also satisfies $e\La H^T = e\La H'^T R^T = s' R^T = s$
so that the answer to {CGW$_{Z}$} is also positive.
We have shown that
{CGW$_Z$} is polynomial-time reducible to {CGW$_{ZF}$}.
By Theorem~\ref{CGW_NpComp},
{CGW$_Z$} is NP-complete and so does {CGW$_{ZF}$}.
} 
%
%
It is trivial that CGW$_{ZF}$ is polynomial-time reducible to QBDD.
Thus by Theorem~\ref{CGW_ZF_NpComp} and by symmetry, we have
\bc \label{Qbdd_NpHard}
QBDD is NP-hard, even if $H=[H_X|O]$ or $H=[O|H_Z]$.
\ec
%
We have shown that QBDD is NP-hard
by considering the practical constraint that the check matrix is of full row rank,
which makes the foundation of
stabilizer code-based cryptography more concrete.
This will also be helpful when we classify of the hardnesses
of the decoding problems
in the next section.

\section{Decoding over the Depolarizing Channel} \label{DecDpCh}

The depolarizing channel is one of the most important channel models
in quantum communication and quantum cryptography
\cite{BS98}, \cite{NC00}, \cite{Smith08}.
In this section,
we will consider the decoding problems
for finding a most likely error
and for finding a most likely error coset
over the depolarizing channel.
Like classical decoding,
it is very intuitive to consider the decoding problem
for finding a most likely error, called
quantum maximum likelihood decoding (QMLD).
But a further analysis shows that
the optimal decoding to minimize the decoding error probability
is to find a most likely error coset,
for which we call it
quantum minimum-error-probability decoding (QMEPD).
We will show that these two problems are NP-hard.

We assume the memoryless model that
the depolarizing channel affects each qubit independently
such that for some $p\in[0, 1]$, a qubit is depolarized
to be a completely mixed state $I/2$ with probability $p$,
and remains intact with probability $1-p$.
%
If $\rho_1$ is a density operator of one-qubit depolarizing channel input,
then it is known that
the channel output can be expressed as
\beqs{
\sE(\rho_1) &=p(I/2)+(1-p)\rho_1\\
&= (1-\veps)\rho_1 + (\veps/3)(X\rho_1 X + Y\rho_1 Y + Z\rho_1 Z)
}
with $\veps = \frac{3}{4}p$.

Now consider an $[[n, k, d]]$ stabilizer code $\sC(\sS)$
with $d\ge 2t+1$ as in Sec.~\ref{StbCode}.
Again the stabilizer group
$\sS=\gen{g_1,g_2,\cdots,g_{n-k}}$
has a check matrix $H$ as in \eqref{ChkMtx}.
Given a coded state $\ket{\psi}\in\sC(\sS)$,
let $\rho=\ketbra{\psi}{\psi}$ be the channel input of the depolarizing channel.
Then the channel output is
\beq{
\sE(\rho) = \sum_{u\in \ZZ_2^{2n}} (\sqrt{\la_u}E_u)\rho(\sqrt{\la_u}E_u)^\dg,
}
where $E_u = \sig_{1} \otimes \sig_{2} \otimes \cdots \otimes \sig_{n}$
with $\sig_j \in \{I,X,Y,Z\}$ such that $\varphi(E_u) = u$
and $\la_u = (\veps/3)^{gw(u)}(1-\veps)^{n-gw(u)}$
since different uses of the channel are independent.
The channel output $\sE(\rho)$ is a mixed state with ensemble
\beql{eq_Ensem}{
\{\la_u, E_u\ket{\psi}\}_{u\in \ZZ_2^{2n}}. 
}
Notice that
a smaller $gw(u)=w(E_u)$ results in a larger $\la_u$
since $0\le \veps = \frac{3}{4}p \le \frac{3}{4}$.

We first consider the decoding problem for finding a most likely error.
Suppose that $\sE(\rho)$ is received
and the $n-k$ syndrome measurements by operators
defined by the generators $g_i$'s of $\sS$ are performed,
as in Sec.~\ref{BdDec}.
Likewise, map the measurement results
to a binary syndrome vector $s\in\ZZ_2^{n-k}$.
%
%
Let $E$ be the unknown channel error operator.
Given the syndrome $s\in\ZZ_2^{n-k}$,
the event $(E = E_u)$ occurs with probability
\beqsl{P(E=E_u|s)}{
& P(E=E_u \mid \text{syndrome is $s$})\\
&= \frac{ P(E=E_u) P(\text{syndrome is $s$} \mid E=E_u) }{ P(\text{syndrome is $s$}) }\\
&= \frac{ \la_u \,  P(\text{syndrome is $s$} \mid E=E_u) }{ q_s } \\
& =\bcase{
  \la_u/q_s  & \text{if $u\La H^T = s$},\\
  0  & \text{otherwise},
  }
}
where
\beqs{
q_s &\teq P(\text{syndrome is $s$})\\
&= \sum_{u\in\ZZ_2^{2n}} P(E=E_u) P(\text{syndrome is $s$} \mid E=E_u) \\
&= \sum_{u\in\ZZ_2^{2n}:\, u\La H^T = s} \la_u
}
is a constant given $s$.
Since a smaller $gw(u)$ results in a larger $\la_u$,
to find a most likely error, we have
\\[8pt]
{\bf \textsc{Quantum Maximum Likelihood Decoding (QMLD)} }\\[2pt]
\textsc{Input:}  A full row rank $m\times 2n$ binary matrix $H$ satisfying $H\La H^T=O$,
                 and a binary vector $s\in\ZZ_2^{m}$.\\[1pt]
\textsc{Output:} A binary vector $e\in\ZZ_2^{2n}$ satisfying $e\La H^T = s$
                 and minimizing $gw(e)$.
\\[8pt]
%
%
Obviously, there is a polynomial-time reduction from QBDD to QMLD.
So by Corollary~\ref{Qbdd_NpHard}, we have
\bc \label{Qmld_NpHard}
QMLD is NP-hard, even if ${H=[H_X|O]}$ or $H=[O|H_Z]$.
\ec

Since QMLD does not limit the search scope of $gw(e)$,
QMLD has better decoding performance than that of QBDD in practice.
But it is known that a decoding rule based on QMLD does not
minimize the decoding error probability \cite{PC08}, \cite{HG10}.
%
To see this,
again let $E$ be the unknown channel error operator.
If we select $E_v^\dg$ as the error-correction operator
for some $v\in \ZZ_2^{2n}$,
then a successful correction will be performed
iff $E_v^\dg (E\rho E^\dg) E_v = \rho$
iff $E_v^\dg E \in \sS\sK$ by \eq{StbProperty}.
Let $P_s(\cdot) \teq P(\,\cdot\, |\, \text{syndrome is $s$})$.
In order to minimize the decoding error probability given $s$,
we have to find a $v\in\ZZ_2^{2n}$ maximizing
\beqs{
& P(E_v^\dg E \in \sS\sK \mid \text{syndrome is $s$})
= P_s(E_v^\dg E \in \sS\sK)\\
&= \sum_{u\in\ZZ_2^{2n}}P_s(E=E_u)\, P_s(E_v^\dg E \in \sS\sK \mid E=E_u) \\
&= \sum_{u\in\ZZ_2^{2n}:\, u\La H^T=s} \frac{\la_u}{q_s}\, P_s(E_v^\dg E \in \sS\sK \mid E=E_u) \quad\text{by \eq{P(E=E_u|s)}} \\
&= \frac{1}{q_s}\sum_{\substack{u\in\ZZ_2^{2n}:\, u\La H^T=s,\\ v+u\in\vph(\sS\sK)=\Row(H)}} \la_u \\
&= \frac{1}{q_s}\sum_{\substack{u\in v+\Row(H):\\ u\La H^T=s}} \la_u \\
&= \bcase{
  \frac{1}{q_s} \sum_{u\in v+\Row(H)} \la_u & \text{if $v\La H^T = s$},\\
  0 & \text{otherwise}.
  }
}
The last equality holds since for all $u\in v+\Row(H)$,
$u\La H^T=s$ iff $v\La H^T = s$ by the fact that $H\La H^T = O$.
Now let $\af_v\teq \sum_{u\in v+\Row(H)} \la_u$ be the aggregate probability of the coset $v+\Row(H)$.
By recalling $\la_u=(\veps/3)^{gw(u)}(1-\veps)^{n-gw(u)}$,
we have
\\[8pt]
{\bf \textsc{Quantum Minimum-Error-Probability Decoding (QMEPD)} }\\[2pt]
\textsc{Input:}  A full row rank $m\times 2n$ binary matrix $H$ satisfying $H\La H^T=O$,
                 a binary vector $s\in\ZZ_2^m$,
                 and a real number $0 \color{red} < \color{black} \veps\le 3/4$.\\[1pt]
\textsc{Output:} A binary vector $v\in\ZZ_2^{2n}$ satisfying $v\La H^T=s$
                 and maximizing $\af_v= \sum_{u\in v+\Row(H)} (\veps/3)^{gw(u)}(1-\veps)^{n-gw(u)}$.
\\[8pt]
%
Note that the optimal decoding problem can be formulated in another way
by assigning each coset $v+\Row(H)$ a unique representative
and limiting the output to be one of those representatives
(see DQMLD in \cite{HG10} or Sec.~IV of \cite{PC08}).
But our formulation for QMEPD is an important step
to classify the complexity of QMEPD.
Now we show
%
\bt \label{Qmepd_NpHard}
QMEPD is NP-hard.
\et
\bPf{
By assuming $H=[H_X|O]$,
we reduce \text{QMLD} to {QMEPD} in polynomial time.
Suppose we have a polynomial-time algorithm for {QMEPD}.
Given an instance of {QMLD}
with some $H=[H_X|O]$ and $s\in\ZZ_2^m$,
let {QMEPD} has the same $H$ and $s$ in its inputs.
Now $|\Row(H)|=2^m$, and observe that
$
\af_v = (1-\veps)^n \sum_{u\in v+\Row(H)}\Big(\frac{\veps/3}{1-\veps}\Big)^{gw(u)}.
$
Let the $\veps$ in {QMEPD} be
sufficiently small such that
$\frac{\veps/3}{1-\veps} < \frac{1}{2^m}$,
and then use the polynomial-time algorithm to solve {QMEPD}.
By our selection of $\veps$,
the algorithm must output a vector $v$
such that the coset $v+\Row(H)$ contains
a solution $e$ to the problem {QMLD}.
Suppose not, i.e., there exists an $e_1\notin v+\Row(H)$ such that
$e_1\La H^T = s$ and $gw(e_1) < gw(e)$ with an $e\in v+\Row(H)$
having a minimum generalized weight among all vectors in the coset $v+\Row(H)$.
Then we have
\beqs{
\af_{e_1}
&= (1-\veps)^n \sum_{u\in e_1+\Row(H)}\Big(\frac{\veps/3}{1-\veps}\Big)^{gw(u)} \\
&\ge (1-\veps)^n \Big(\frac{\veps/3}{1-\veps}\Big)^{gw(e_1)} \\
&\ge  (1-\veps)^n \Big(\frac{\veps/3}{1-\veps}\Big)^{gw(e)-1} \\
&> (1-\veps)^n 2^m \Big(\frac{\veps/3}{1-\veps}\Big)^{gw(e)} \\
&\ge (1-\veps)^n \sum_{u\in v+\Row(H)}\Big(\frac{\veps/3}{1-\veps}\Big)^{gw(u)} \\
&= \af_v,
}
a contradiction to the maximality of $\af_v$.
However, we only have $v=(x|z)$ for some $x,z\in\ZZ_2^n$.
To obtain a solution to {QMLD} from $v$,
let $e'\teq ({\bf 0}|z)\in\ZZ_2^{2n}$.
First, $e'$ satisfies $e'\La H^T=s$ since $H=[H_X|O].$
That means  $gw(e')\ge gw(e)$ by the minimum of the $gw(e)$ in {QMLD}.
But $e\in v+\Row(H)$ implies
$e=(x+h|z)$ for some $h\in\Row(H_X)$,
so $gw(e)\ge w_H(z) = gw(e')$ by \eq{gw_ge_wH}.
We have shown that $e'\La H^T=s$ and $gw(e')=gw(e)$,
i.e, the vector $e'$ constructed from $v$
is also a solution to {QMLD}.
Thus {QMLD} is polynomial-time reducible to {QMEPD},
as $H=[H_X|O]$ is assumed.
By Corollary~\ref{Qmld_NpHard}, {QMEPD} is NP-hard.
} 
%
%
A trick of the proof is to set
a sufficiently small $\veps$ such that $\frac{\veps/3}{1-\veps} < \frac{1}{2^m}$.
One may argue that, in practice,
the channel may have a larger $\veps$, which may make the decoding easier.
But if there exists a decoder that can work efficiently
over a depolarizing channel with some channel parameter $\veps_1$,
then the decoder is expected to work efficiently
over a depolarizing channel with a smaller channel parameter $\veps_2 < \veps_1$.
Then the proof above suggests that
the decoder should have huge complexity unless P=NP.
%
%
Also note that in order to perform QMEPD practically,
an auxiliary channel estimation may be needed
to estimate the actual $\veps$ of the channel,
and to compute $\af_v$, the exponential function
needs large space complexity \cite{CC88}, \cite{Furer07}.
However, 
for the time complexity,
the NP-hardness of QMEPD in Theorem~\ref{Qmepd_NpHard}
answers the dangling problem of
how hard an optimal decoding over the depolarizing channel is.

In quantum cryptography,
the hardnesses of QMLD and QMEPD
reflect the hardness of eavesdropping on
a stabilizer code-based cryptosystem
over the depolarizing channel model,
i.e., the artificial noise used to protect the system
is generated in some way like the depolarizing channel.
Corollary~\ref{Qmld_NpHard} and Theorem~\ref{Qmepd_NpHard}
indicate that such a system can effectively resist
the attacks based on QMLD or QMEPD.

\section{Conclusion} \label{Summary}

The complexities of QBDD, QMLD and QMEPD
are classified in this paper.
The weight metric used is the generalized weight and
the check matrices in the decoding problems are of full row rank.
In this paper, QBDD is shown to be NP-hard,
regardless of any specific channel model considered.
Then over the depolarizing channel,
both QMLD and QMEPD are shown to be
NP-hard by showing that
there are polynomial-time reductions
from QBDD to QMLD and
from QMLD to QMEPD.
The NP-hardnesses of these decoding problems suggest that
decoding general stabilizer codes is extremely difficult.
But this decoding difficulty strengthens the foundation of quantum code-based cryptography.


%


\end{document}